\documentclass[useAMS,usenatbib,usegraphicx]{mn2e}
\usepackage{lineno}
\usepackage{xcolor}
\usepackage{pdflscape}

\newcommand{\etal}{{et al.\ }}
\newcommand{\be}{\begin{equation}}
\newcommand{\ee}{\end{equation}}
\newcommand {\apgt} {\ {\raise-.5ex\hbox{$\buildrel>\over\sim$}}\ }
\newcommand {\aplt} {\ {\raise-.5ex\hbox{$\buildrel<\over\sim$}}\ }

\title[Vetting $Kepler$ planet candidates]{\textcolor{black}{Vetting $Kepler$ planet candidates in the sub-Jovian desert with multi-band photometry}}

\author[K. D. Col\'on \etal]{Knicole D. Col\'on$^{1}$\thanks{E-mail: kcolon@lehigh.edu}, Robert C. Morehead$^{2,3}$\thanks{NSF Graduate Research Fellow.}, Eric B. Ford$^{2,3}$\\
$^{1}$Department of Physics, Lehigh University, 16 Memorial Drive East, Bethlehem, PA 18015, USA \\
$^{2}$Department of Astronomy and Astrophysics, The Pennsylvania State University, 525 Davey Laboratory, State College, PA 16803, USA \\
$^{3}$Center for Exoplanets \& Habitable Worlds, The Pennsylvania State University, 525 Davey Laboratory, State College, PA 16803, USA}

\begin{document}

\date{Accepted....  Received...; in original form 2014 September}

\pagerange{\pageref{firstpage}--\pageref{lastpage}} \pubyear{2015}

\maketitle

\label{firstpage}
 
\begin{abstract} 
\textcolor{black}{We present new multi-band transit photometry of three small ($R_{p} \aplt 6$ $R_{\oplus}$), short-period ($P \aplt 6$ days) $Kepler$ planet candidates acquired with the Gran Telescopio Canarias.}  These observations \textcolor{black}{supplement} the results presented in \citet{colon11} and \citet{colon12}, where we used multicolor transit photometry of five $Kepler$ planet candidates to search for wavelength-dependent transit depths and \textcolor{black}{either validate planet candidates or} identify eclipsing binary false positives within our sample.  In those previous studies, \textcolor{black}{we provided evidence that three targets were false positives and two targets were planets}.  \textcolor{black}{Here, we present observations that provide evidence supporting a planetary nature for KOI 439.01 and KOI 732.01,} \textcolor{black}{and we find that KOI 531.01, a 6 $R_{\oplus}$ planet candidate around an M dwarf, is likely a false positive.}  We also present a discussion of the purported ``sub-Jovian desert'' in the orbital period-planet radius plane, \textcolor{black}{which cannot be easily explained by observational bias}.  \textcolor{black}{Both KOI 439.01 and KOI 732.01 are \textcolor{black}{likely planets} located within the so-called desert and should be investigated with further follow-up observations.}  As \textcolor{black}{only $\sim$30 of the $\sim$3600 currently active $Kepler$ planet candidates are} located within the sub-Jovian desert, it will be interesting to see if these candidates also survive the vetting process and fill in the gap in the period-radius plane.  \textcolor{black}{Confirming planets in this regime will be important for understanding planetary migration and evolution processes, and we urge additional follow-up observations of these planet candidates to confirm their nature.}
\end{abstract}

\section{Introduction} 
\label{sintro}

With the discovery of nearly 1000 planets and several thousand additional planet candidates awaiting confirmation,\footnote{Up to date numbers can be found at http://kepler.nasa.gov/} the $Kepler$ space mission has set the stage for in-depth studies of different planet populations.  In order to study these populations, the reliability of the $Kepler$ sample must be taken into account.  \citet{morton11}, \citet{morton12}, \citet{fressin13}, and \citet{desert12} find a low global false positive rate for $Kepler$ candidates of $\sim$10\%, suggesting that most of the remaining candidates are real planets.  However, \citet{coughlin14} estimate a 35\% false positive rate for $Kepler$ candidates due to various types of contamination.  \citet{santerne12} and \citet{colon12} also suggest higher false positive rates of 35\% and $\sim$50$-$60\% for specific subsets of $Kepler$ candidates, such as small candidates in single systems with orbital periods approximately less than six days \citep{colon12}.  On the other hand, nearly all candidates in multi-candidate systems are expected to be real \citep{liss14,rowe14}.

Given this range of false positive rates, it is unclear how significantly false positives affect statistical studies of different subsets of $Kepler$ candidates (e.g. \citealp{howard12}).  In particular, there is an apparent excess of single-candidate systems (relative to multi-candidate systems), but it remains to be seen what fraction of the single-candidate systems are real.  For this reason, candidates believed to be in single systems are the focus of this paper.  Such candidates with short-periods are particularly interesting, given that some are located within an apparent sub-Jovian desert in the orbital period-planet radius plane \citep{sk11,bn13}.

In \citet{colon11} and \citet{colon12} (hereafter, Paper I and Paper II), we presented multi-wavelength transit observations of five small, short-period $Kepler$ planet candidates (KOI 565.01 and KOI 225.01, 420.01, 526.01, and 1187.01) acquired with the 10.4-m Gran Telescopio Canarias (GTC).  A complete description of our target selection criteria is given in Paper II and is summarized here, again with only single-candidate systems being considered:
\begin{itemize}
\item orbital period $(P) \aplt 6$ days
\item planet radius $(R_{p}) \aplt 6$ $R_{\oplus}$
\item transit depth ($\delta$) $\apgt$ 500 ppm
\item transit duration ($\tau$) $\aplt$ 2.5 h
\item $Kepler$ magnitude (Kp) $\aplt$ 15.5
\end{itemize}
\textcolor{black}{Currently, there are 56 active KOIs that satisfy these criteria, including KOI 420.01 and 526.01 (the two systems we validated as planets in Paper II).}  In these previous papers, we looked for wavelength-dependent transit depths for each target, which is suggestive of an eclipsing binary false positive scenario assuming the target is not a system composed of two (or more) equal mass or equal temperature eclipsing stars (which would reveal no color difference).  Of five targets observed, we identified three as having achromatic light curves and validated two as real planets.  The resulting high false positive rate for our sample was not surprising, given the large percentage of eclipsing binaries with short orbital periods that can mimic planetary transits (particularly when diluted by other nearby or unresolved stars). \textcolor{black}{We note that we did not inspect the $Kepler$ light curves of our targets in detail prior to observing them.  We assumed the targets had been appropriately vetted prior to being added to the candidate list, so we did not search for secondary eclipses in the $Kepler$ data nor did we identify candidates with particularly v-shaped light curves.  However, targets displaying such attributes would be more likely to be false positives than a typical candidate, so we encourage similar studies to perform detailed inspections of $Kepler$ light curves prior to collecting any follow-up data.}  

In this paper, we present multi-wavelength transit observations from the GTC of three additional small, short-period $Kepler$ planet candidates (KOI 439.01, 531.01, and 732.01).  Tables \ref{props} and \ref{cprops} list some relevant properties for each of these targets.  In Sections \ref{sobs} and \ref{sdata}, we present a summary of our observations and our data reduction and analysis.  Results for each target are presented in Section \ref{sresults}.  A discussion of our results for each target, as well as an updated investigation of trends in the false positive rate with different planetary and stellar parameters, is given in Section \ref{sdiscuss}.  We also include a discussion of candidates located within the apparent sub-Jovian desert in Section \ref{sdiscuss}, and we offer concluding remarks in Section \ref{sconc}.

\begin{table}
   \caption{KOI Star Properties \label{props}}
  \begin{tabular}{@{}ccccc@{}}
  \hline
KOI & KIC & Kp & T$_{eff}$ (K) & $b$ (deg) \\
\hline
439   & 12470954 & 14.313 & 5438 & 13.06 \\ 
531   & 10395543 & 14.418 & 4030 & 16.61 \\
732   & 10265898 & 15.342 & 5379 & 16.15 \\
\hline
\end{tabular}

\medskip
All values are from the NASA Exoplanet Archive (http://exoplanetarchive.ipac.caltech.edu/; accessed on 2015 April 4).  The KIC number is the $Kepler$ Input Catalog number for each target.  Here, $b$ is the Galactic latitude of the KOI host star.

\end{table}


\begin{table*}
 \centering
 \begin{minipage}{170mm}
  \caption{KOI Candidate Properties \label{cprops}}
  \begin{tabular}{@{}cccccccc@{}}
  \hline
KOI & $P$ (days) & $\tau$ (d) & $b$ & $R_{p}/R_{\star}$ & $R_{p}$ ($R_{\oplus}$) & $\delta$ (ppm) & EB Catalog \\
\hline
439.01   & 1.90221$\pm$4.1e-07 & 0.09199$\pm$2.875e-04 & 0.0433$^{+1.69e-01}_{-4.32e-02}$ & 0.0435$^{+3.04e-04}_{-9.40e-05}$ & 3.92$^{+1.19e-00}_{-2.50e-01}$ & 2341$\pm$7.900e+00 & yes  \\ 

531.01   & 3.68747$\pm$3.9e-07 & 0.04198$\pm$7.500e-04 & 0.907$^{+2.32e-02}_{-1.09e-02}$ & 0.0868$^{+6.43e-03}_{-3.31e-03}$ & 5.69$\pm$6.60e-01 & 5295$\pm$2.550e+01 & yes \\

732.01   & 1.26026$\pm$6.6e-07 & 0.07809$\pm$6.542e-04 & 0.788$^{+1.25e-02}_{-5.53e-01}$ & 0.0353$^{+3.80e-05}_{-3.11e-03}$  & 3.69$^{+1.30e-00}_{-2.90e-01}$  & 1207$\pm$1.030e+01 & no \\
\hline
\end{tabular}

\medskip
All values are from the NASA Exoplanet Archive (http://exoplanetarchive.ipac.caltech.edu/; accessed on 2015 April 4).  Here, $\tau$ is the transit duration, $b$ is the impact parameter, and $\delta$ is the transit depth as measured in the $Kepler$ bandpass.  The last column indicates if the KOI is listed in the third version of the $Kepler$ eclipsing binary catalog (\citealp{slawson11}; http://keplerebs.villanova.edu/; accessed on 2015 April 4).

\end{minipage}
\end{table*}


\section{Observations}
\label{sobs}

We used OSIRIS on the GTC to acquire near-simultaneous multicolor transit photometry of three $Kepler$ planet candidates: KOI 439.01, KOI 531.01, and KOI 732.01.  This facility is described in further detail in Papers I and II.  As in Paper II, we alternated between two broadband order sorter filters during each transit observation: 666 $\pm$ 36 nm and 858 $\pm$ 58 nm.  All observations were conducted in queue (service) mode and used 1 $\times$ 1 binning and a fast readout rate of 500 kHz.  We describe specific details for each observation in the following sections.  

\subsection{KOI 439.01}
\label{sobs1}

The 2012 June 11 \textsc{ut} transit of KOI 439.01 was observed under photometric conditions during gray time.  Observations began at 00:18 \textsc{ut} and ended at 04:30 \textsc{ut} on 2012 June 11, during which time the airmass varied from 1.08 to 1.43.  At 01:40 \textsc{ut} there was a technical problem with the primary mirror, and the time series was interrupted.  Observations resumed at 02:04 \textsc{ut}.  The seeing was extremely variable in the first part of the night, ranging from 0.9 to 1.7 arcsec \textcolor{black}{(7.1 to 13.4 pixels)}.  A slight defocus was implemented after 03:00 \textsc{ut} due to an overall improvement in the seeing.  The integration time was set to 60 s for both filters, and a single window of 1900 $\times$ 2100 pixels located on one CCD chip was read out.  The resulting dead time between each exposure was 21 s.  Autoguiding kept the target centroid stable to within 7 pixels. 

\subsection{KOI 531.01}
\label{sobs2}

We observed the 2012 September 5 \textsc{ut} transit of KOI 531.01 in photometric/clear conditions during bright time.  The airmass ranged from 1.07 to 1.53 during the observations, which took place from 22:01 \textsc{ut} (2012 September 5) to 01:20 \textsc{ut} (2012 September 6).  The seeing was unstable (varying between 0.8 and 1.5 arcsec \textcolor{black}{or 6.3 and 11.8 pixels}), so the focus was adjusted several times to avoid saturating the target star.  Still, the target was saturated in four images, and we exclude these from our analysis.  The integration time used for both filters was 23 s.  The full CCD chip where the target was located was read out, resulting in a dead time between exposures of 29.5 s.  The target centroid drifted by $<$ 4 pixels during the observations.  

\subsection{KOI 732.01}
\label{sobs3}

Observations of the 2012 May 27 \textsc{ut} transit of KOI 732.01 took place under clear conditions and during gray time.  The observations began at 00:42 \textsc{ut} and ended at 05:12 \textsc{ut} on 2012 May 27.  In that time the airmass ranged from 1.05 to 1.37 and the seeing varied between 0.8 and 1.0 arcsec \textcolor{black}{(6.3 and 7.9 pixels)}.  No defocus was applied.  An integration time of 60 s was used for both filters.  Both CCD chips were read out, but in our analysis we only use comparison stars that are located on the same chip as the target to minimize any systematic differences in the properties of the two chips.  The corresponding dead time between exposures was 34 s.  Autoguiding kept the target centroid stable to within 5 pixels, but there are a few outliers towards the end of the observations where the centroid drifted by as much as 11 pixels.

\section{Data Reduction and Analysis}
\label{sdata}

All images were reduced using software written in GDL.\footnote{GNU Data Language; http://gnudatalanguage.sourceforge.net/.}  We performed bias subtraction and flat fielding on each science image prior to performing aperture photometry.  Dome flats acquired for each filter were used, and we corrected for non-uniform illumination by the dome lamp by removing the large-scale illumination pattern in the final co-added flat field for each bandpass through smoothing.  We note that no flats were taken for the 858/58 filter for the KOI 531 observations.  We therefore use the 858/58 flats from the KOI 732 observations in our analysis of the KOI 531 data, as both observations utilized the full CCD chip.  Although these two observations were taken about three and a half months apart, we do not believe our analysis was significantly affected by using ``old'' flats on the KOI 531 858/58 data.  In particular, we find that the standard deviation of the time-binned residuals for the KOI 531 light curve at 858-nm is consistent with the trend expected for white Gaussian noise (despite systematics that are clearly seen in the light curves; \S\ref{sresults}).
 
Circular aperture photometry was performed on each target, stars within $\sim$20 arcsec of the target (i.e. nearby stars that could be the source of the transit signal), and a sample of comparison stars.  We initially used apertures with a range of radii, but ultimately the aperture that yielded the best photometry (i.e. the lowest rms scatter outside of the transit) for each target was used.  For KOI 439, an aperture of 24 and 26 pixels (3.0 and 3.3 arcsec) was used for the 666-nm and 858-nm data, respectively, and an annulus with an inner radius of 60 pixels and outer radius of 65 pixels (7.6 and 8.2 arcsec) was used for sky subtraction.  For KOI 531, an aperture of 26 and 24 pixels (3.3 and 3.0 arcsec) was used for the 666-nm and 858-nm data, respectively, and an annulus with inner and outer radii of 50 and 55 pixels (6.3 and 7.0 arcsec) was used for sky subtraction.  For KOI 732, an aperture of 25 and 34 pixels (3.2 and 4.3 arcsec) was used for the 666-nm and 858-nm data, respectively, and an annulus with inner and outer radii of 70 and 75 pixels (8.9 and 9.5 arcsec) was used for sky subtraction.  To avoid including the target flux in the sky annulus of the nearby stars we checked as potential sources of the transit signal, we sometimes applied a larger sky annulus for those stars.        

We generated a light curve for each target by dividing the total flux measured from the target star by the total weighted flux of an ensemble of comparison stars.  Comparison stars that had a similar brightness to the target and that did not display variability were included in the ensemble.  For consistency, for a given target the same ensemble was used to generate the 666- and 858-nm light curves.  For KOI 439, 531, and 732, ensembles of three comparison stars were ultimately used.  We excluded any data where the peak counts in the target were $>$ 50000 to avoid the saturation limit and stay in the linear regime of the CCD.  The light curves were then normalized by dividing by the median flux ratio measured in the out-of-transit data.  

The observation epochs at mid-exposure were extracted from the FITS header for each image.  The extracted \textsc{UTC} times were converted to Barycentric Julian Dates in Barycentric Dynamical Time (BJD\_TDB) via an online applet\footnote{http://astroutils.astronomy.ohio-state.edu/time/utc2bjd.html; \citet{eastman10}} for consistency with the time coordinate system that the $Kepler$ mission uses.

The out-of-transit light curve for each target was then regressed against airmass, the target centroid position, and the peak counts in the target (per pixel), and a linear trend was removed from the full light curve for each target.  For KOI 531, we also linearly detrended the light curve against the sky background and removed a quadratic trend that was apparent in the baseline (out-of-transit) data.

The median photometric error for the target light curve, which was computed from the photon noise of the target and comparison stars, the noise from the sky background, and scintillation, is 577 and 522 ppm for the 666- and 858-nm light curves for KOI 439, 838 and 682 ppm for the 666- and 858-nm light curves for KOI 531, and 693 and 613 ppm for the 666- and 858-nm light curves for KOI 732.  In all cases, the photon noise of the target and reference star ensembles dominates the error budget.  In most cases, the standard deviation of the light curve residuals is larger than the computed photometric error, \textcolor{black}{suggesting that our computed errors may be underestimated}.  We therefore perform a prayer-bead analysis to better characterize the uncertainties on each fitted parameter in our light curve model.  We discuss this further below.

We followed a similar procedure as in Paper II to fit a synthetic limb-darkened light curve model \citep{mandel02} to the detrended data for each target.  The process is summarized here:  For each target, we fitted models to each light curve separately and then corrected the data against the best-fitting (linear) baseline slope.  Light curve residuals were computed by subtracting the model from the data and were used to identify and discard outlying data points (i.e. points greater than 3$\sigma$ from the median value of the residual light curve).  We then fitted models to the corrected light curves simultaneously and forced the mid-transit time ($t_c$), transit duration (from first to fourth contact; $\tau$), impact parameter ($b$), baseline flux ratio, and baseline slope to the same value for both light curves.  The planet-star radius ratio ($p$ = $R_p$/$R_{\star}$) and limb-darkening coefficients were determined separately for each wavelength.  \textcolor{black}{The fitting process involved iterating over a range of initial guesses for each parameter.  In each iteration, each limb-darkening coefficient was held fixed at a specific value for each wavelength rather than being allowed to vary freely like the other model parameters.  However, each iteration used a different set of values for the limb-darkening coefficients to account for uncertainties in the coefficients.}  A prayer-bead analysis was also performed here, where we fitted models to synthetic data generated from circularly shifting the residuals of each light curve.  In Section \ref{sresults}, we consider the distribution of the best-fitting parameters to the real and synthetic data when discussing the uncertainties for each fitted parameter in order to account for any sources of systematic noise.    

Finally, we computed the false positive probability (FPP) for each KOI presented in this paper using $vespa$.  This tool is based on the Morton (2012) procedure for calculating FPPs for transiting planet candidates discovered by the $Kepler$ mission and was recently made available for general use by the community.\footnote{http://github.com/timothydmorton/vespa}  The $vespa$ tool computes FPPs by simulating three different astrophysical scenarios that could mimic a planetary transit: (1) an undiluted eclipsing binary (EB), a hierarchical eclipsing binary system (HEB), or a background (or foreground) eclipsing binary blended with the target (BEB).  \textcolor{black}{We specifically used $vespa$ to calculate FPPs for each of our KOIs from the $Kepler$ photometry (downloaded from the NASA Exoplanet Archive), inputting the right ascension and declination, orbital period, planet-star radius ratio (measured from $Kepler$; Table \ref{cprops}),  and the Kp, $V$, $griz$, and $JHK$ magnitudes for each KOI.  Given the lower photometric precisions achieved in the GTC observations presented here, we only compute FPPs based on the $Kepler$ photometry.  The results of our analysis are presented in the following sections.}
 

\section{Results}
\label{sresults}

\begin{figure}
\includegraphics[scale=0.3, angle=90]{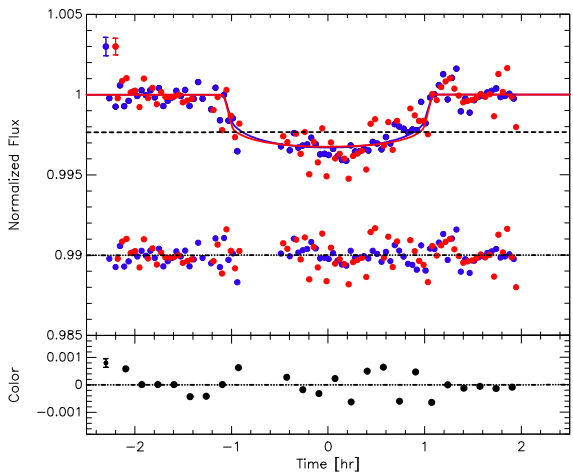}
\caption{Transit light curves, residuals, and color for the 2012 June 11 \textsc{UT} transit of KOI 439.01.  The blue and red points represent the 666 nm and 858 nm data, respectively.  The solid curves are the best-fit models.  Representative photometric error bars are shown on the upper left-hand side of the top and bottom panels.  The dashed line indicates the depth of the transit  measured in the $Kepler$ bandpass (Table \ref{cprops}).  The dash-dot line in the top panel illustrates the quality of the light curve model.  The dash-dot line in the bottom panel indicates where the color is zero (for reference). \textcolor{black}{The errors on the color are significantly smaller than the individual photometric errors shown in the top panel as a result of binning the data prior to computing the color.}}
\label{lc439}
\end{figure}

\begin{figure}
\includegraphics[scale=0.3, angle=90]{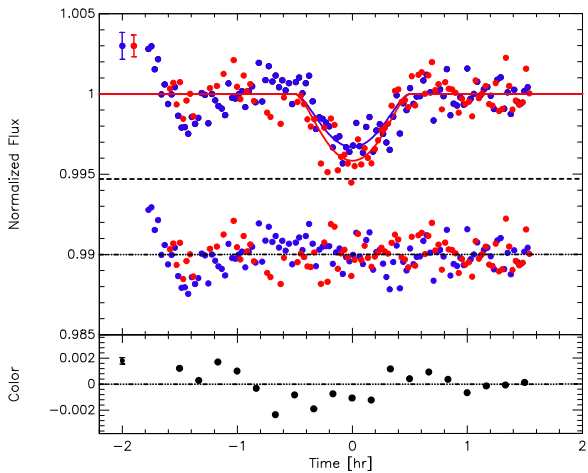}
\caption{Same as Figure \ref{lc439}, but for the 2012 September 5 \textsc{UT} transit of KOI 531.01.}
\label{lc531}
\end{figure}

\begin{figure}
\includegraphics[scale=0.3, angle=90]{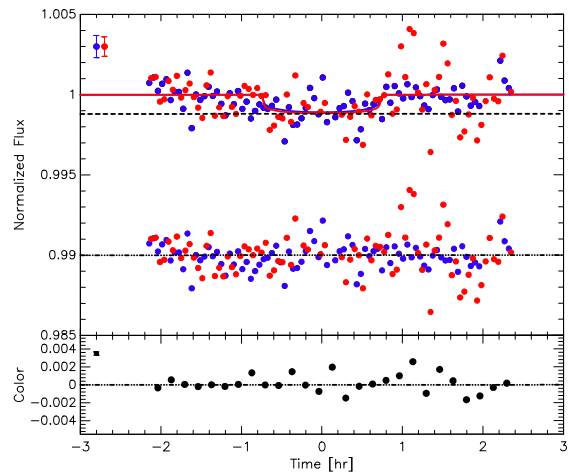}
\caption{Same as Figure \ref{lc439}, but for the 2012 May 27 \textsc{UT} transit of KOI 732.01.}
\label{lc732}
\end{figure}

In Figs \ref{lc439}$-$\ref{lc732} we present the light curves and the best-fitting models for each target, along with the light curve residuals and the color (666$-$858 nm).  \textcolor{black}{To compute the color we binned each light curve in 10-minute intervals and then computed the color as -2.5 $log$($F_{666}$/$F_{858}$).  Here, $F_{666}$ is the flux ratio measured in the 666 nm bandpass and $F_{858}$ is that measured in the 858 nm bandpass}.  Thus in Figs \ref{lc439}$-$\ref{lc732}, a positive color indicates a `red' transit and a negative color indicates a `blue' transit.    

While not presented here, we also examined the light curves for stars within 20 arcsec of each target to confirm that the target is the source of the transit signal.  We found no transit-like signals in any nearby stars, confirming that in each case the target is \textcolor{black}{most likely} the variable star.  However, we cannot rule out that the signal is due to an object transiting an unresolved star (specifically within 2.5 arcsec of the target for KOI 439.01 and KOI 531.01; \S\ref{sres1} and \S\ref{sres2}). 

In Table \ref{fits} we present the best-fitting model parameters for each target.  \textcolor{black}{The best-fitting parameters are identified as the median best-fitting value from the prayer-bead chain.  The 1$\sigma$ uncertainties for each parameter are also presented in Table \ref{fits}, which are computed as the standard deviation of each parameter fitted in the prayer-bead analysis.  We also present in Table \ref{rrr} the median ratio of the planet-star radius ratios measured at 666-nm and 858-nm (i.e. $p_{666}/p_{858}$).}

\textcolor{black}{In Table \ref{fpp} we present the results from the FPP calculations.  The table contains the relative probability that the measured light curve is either a planetary transit, an EB, a HEB, or a BEB.  The final column contains the FPP.}  

Results for each individual target are summarized here and are discussed in further detail in the following section.

\begin{description}

\item[\emph{KOI 439.01}:] We find that the transit depths in the two GTC bandpasses are consistent,\footnote{While the depth in the $Kepler$ bandpass is shallower than in the GTC bandpasses, we attribute this to dilution resulting from a nearby star that contaminated the $Kepler$ aperture.} and the colour shows no significant change during the transit.  The measured planet-star radius ratios for the two bandpasses are also consistent \textcolor{black}{within 3$\sigma$ (Table \ref{rrr})}.  \textcolor{black}{The FPP computed from the $Kepler$ light curve is 1.91\%.  Given the low FPP from $Kepler$ and that we measure no significant difference in the planet-star radius ratios, we believe this candidate is likely a planet.  However, we encourage additional follow-up to confirm these findings.  We discuss this target further in \S\ref{sres1}.}  

\item[\emph{KOI 531.01}:]  \textcolor{black}{Systematics are clearly present in the light curve prior to the transit, but we find no correlation between the systematic trends seen and either instrumental or astrophysical parameters.}  \textcolor{black}{Despite the systematics, we find that the planet-star radius ratios differ in the two bandpasses by $>$3$\sigma$ (Table \ref{rrr}), suggesting this to be a false positive.  \textcolor{black}{Also, the FPP we compute from the $Kepler$ light curve is very high (90.0\%).  However, we only measure a $1.5\sigma$ difference in the colour during transit compared to out-of-transit, which makes the identity of this target somewhat ambiguous}.  We discuss these results further in \S\ref{sres2}.}

\item[\emph{KOI 732.01}:]  The light curve suffers from low signal-to-noise due to the faintness of the target (Kp = 15.3), but we find no visual evidence of a wavelength-dependent transit depth.  \textcolor{black}{However, the measured planet-star radius ratios differ by $>$3$\sigma$ (Table \ref{rrr}).  We note that the scatter in the color measured for KOI 732 is significantly smaller than the scatter for our other targets.  This could be due to particularly favorable observing conditions and/or unusually low stellar activity for KOI 732.  Given the relatively small number of observations used to estimate the photometric precision, we are cautious not to over-interpret these results.  We recognize that the scatter of photometry might be smaller than expected given the relatively small number of observations.} \textcolor{black}{Furthermore, the low FPP of 0.544\% measured from the $Kepler$ data support our visual findings that there is no significant depth difference between our bandpasses.  Overall, we conclude that this is likely a real planet }(which is further supported by the multiplicity of this system; \S\ref{sres3}).

\end{description}

\begin{table*}
\centering
 \begin{minipage}{250mm}
  \caption{Best-Fit Model Parameters \label{fits}}
  \begin{tabular}{@{}cccccccccc@{}}
  \hline
KOI & $t_0$ & $\tau$ & $b$ & $R_{p}/R_{\star}$ & $R_{p}/R_{\star}$ & $c_1$ & $c_2$  & $c_1$ & $c_2$ \\
 & (BJD\_TDB$-$2455000) & (d) &  & (666 nm)  & (858 nm) & (666 nm) & (666 nm) & (858 nm) & (858 nm) \\
\hline
439.01 &  1089.60915$\pm$2.0796e-04  &  0.0912$\pm$1.30e-03  &  0.0000$\pm$7.231e-02  &  0.0513$\pm$1.18e-03  &  0.0526$\pm$8.45e-05  &    0.7099  &    0.3153  &    0.5536  &   0.1330  \\
531.01 &  1176.49398$\pm$8.6418e-06  &  0.0413$\pm$1.29e-04  &  0.9675$\pm$2.133e-03  &  0.0809$\pm$1.01e-03  &  0.0872$\pm$9.33e-04  &    0.7212  &   -0.0038  &    0.6056  &  -0.2044  \\
732.01 &  1074.62127$\pm$3.8018e-05  &  0.0630$\pm$1.69e-04  &  0.0000$\pm$4.507e-03  &  0.0300$\pm$2.25e-05  &  0.0318$\pm$3.95e-05  &    0.6887  &    0.1917  &    0.5583  &  -0.0913  \\
\hline
\end{tabular}
\end{minipage}
\end{table*}


\begin{table}
   \caption{Median Ratios of Planet-Star Radius Ratios \label{rrr}}
  \begin{tabular}{@{}ccc@{}}
  \hline
KOI & $p_{666}/p_{858}$ & $\sigma_{p_{666}/p_{858}}$$^a$ \\
 \hline
439.01 & 0.975 & 0.082 \\
531.01 & 0.928 & 0.046 \\
732.01 & 0.943 & 0.004 \\
\hline
\end{tabular}

\medskip
\textcolor{black}{$^{a}$ The 3$\sigma$ value is presented here.}  

\end{table}


{\begin{table*}
 \centering
 \begin{minipage}{180mm}
   \caption{False Positive Probabilities \label{fpp}}
  \begin{tabular}{@{}ccccccc@{}}
  \hline
KOI & $\lambda_c$ (nm) & $P_{PL}$ & $P_{EB}$ & $P_{HEB}$ & $P_{BEB}$ & FPP \\
 \hline
439.01 & Kepler & 0.981 & 2.94e-5 & 5.56e-13 & 0.0191 & 0.0191 \\
531.01 & Kepler & 0.100 & 0.857 & 1.87e-4 & 0.0428 & 0.900 \\
732.01 & Kepler & 0.995 & 2.72e-15 & 6.33e-34 & 5.44e-3 & 5.44e-3  \\
\hline
\end{tabular}

\medskip

\end{minipage}
\end{table*}


%
%
%

\section{Discussion} 
\label{sdiscuss}

\subsection{KOI 439.01}
\label{sres1}

\citet{lillo12} identified a companion 5.34 mag fainter in $i$-band and 5.453 arcsec from the target.  As our aperture was 3.0-3.3 arcsec, we avoided contamination from this companion.  \citet{law13} also ruled out companions within 2.5 arcsec of the target.  From this, we believe that the transit signal is indeed coming from the target.  \textcolor{black}{Our FPP calculations support this as well (Table \ref{fpp} and \S\ref{sresults}).  However, we recommend radial velocity as well as additional photometric follow-up to help confirm the nature of this target.  Such observations could allow a background eclipsing binary scenario to be confidently ruled out.}

It should be noted that KOI 439.01 was included in the third version of the $Kepler$ eclipsing binary catalog.\footnote{However, it is our understanding that the third version of the eclipsing binary catalog includes all $possible$ eclipsing binary stars, and it is not a comprehensive catalog of $confirmed$ $Kepler$ eclipsing binaries; http://keplerebs.villanova.edu/}  Furthermore, a second planet candidate was found in this system with a period of 5.4 days and a depth of 117 ppm (KOI 439.02), but was later flagged as a false alarm \citep{liss14,rowe14}.  While unrelated, these facts do raise a cautionary flag towards the planetary nature of KOI 439.01.  Regardless, our observations show no \textcolor{black}{obvious evidence that this target is an eclipsing binary false positive or false alarm, strengthening the case that KOI 439.01 is likely a planet}.

This target was noted by \citet{bn13} as a particularly interesting planet, if confirmed.  This is because based on the estimated planet radius (3.9 $R_{\oplus}$) and the short orbital period (1.9 days), this candidate sits within what is becoming known as the sub-Jovian desert.  We discuss this region in further detail in \S\ref{pampas}.

\subsection{KOI 531.01}
\label{sres2}

\textcolor{black}{This candidate has been of particular interest to groups studying planets around cool stars, as the estimated stellar effective temperature is $\sim$4000 K \citep{muirhead12,dressing13,morton14,swift15}.  While commonly treated as a planet, \citet{morton14} notably find KOI 531.01 to have a calculated false positive probability of 99\%.  A more recent calculation using all available quarters of $Kepler$ data yielded a false positive probability of 48\% \citep{swift15}, which is still high but not conclusive evidence of a false positive nature.  Furthermore, this target was noted in \citet{borucki11} as having a ``strange'' light curve.  The phase-folded $Kepler$ light curve presented in the data validation summary from the NASA Exoplanet Archive shows a clear and fairly v-shaped transit.  The ``strangeness'' appears to come from several points during transit that appear to be mis-phased and possibly suggest the presence of a secondary eclipse, as they are shallower than other points during transit.  However, the latest light curve generated using data from Q1-Q17 shows no outliers and no sign of a secondary event.  Thus, the only published observational evidence of this candidate being a false positive is from \citet{swift13}, who claim to have found a deep secondary eclipse for KOI 531.01 in the $Kepler$ data (which appears to contradict the latest light curve from the $Kepler$ team).  While the raw $Kepler$ light curve may hint at a potential depth difference in odd vs. even transits, it is not definitive.  Furthermore, while KOI 531.01 is also included in the latest $Kepler$ eclipsing binary catalog, the light curve in the catalog is not particularly informative and does not display evidence of a secondary eclipse or depth differences.  Lastly, some evidence of transit-timing variations was found by \citet{ford11}, but it was noted that the short duration likely affected the transit times so the variations may not be significant.}   

\textcolor{black}{The depth difference measured in our photometry (Table \ref{rrr}) combined with the high FPP computed by our group (and by \citealp{morton14} and \citealp{swift15}) from the $Kepler$ photometry} suggests that KOI 531.01 is indeed a false positive.  Our FPP calculations specifically suggest that KOI 531.01 is itself an EB (rather than being a hierarchical system or a blend with a background eclipsing system). This is supported by \citet{law13}, who detected no companions down to $\Delta$$m$ = 5 within 2.5 arcsec.  Given that \citet{law13} was sensitive only to targets up to $\sim$5 mag fainter at distances of 0.5 arcsec from the target, it is still possible that there is a faint stellar companion blended with the host star that either affected our photometry and/or is the source of the observed transit signal.  \textcolor{black}{Ultimately, because the measured color difference has only a marginal significance (1.5$\sigma$), we declare that the nature of KOI 531.01 remains somewhat ambiguous.}   We urge radial velocity and additional photometric follow-up \textcolor{black}{(especially at redder wavelengths)} to help clarify the nature of this system.  If real, it would be one of the few known giant ($\sim$6 $R_{\oplus}$) planets with an M dwarf host.

\subsection{KOI 732.01}
\label{sres3}

After our observations were conducted, the $Kepler$ team identified two additional candidate planets around KOI 732.  Given the extremely low false positive probability for candidates in multi-planet systems \citep{liss14,rowe14}, \textcolor{black}{this planet has already effectively been validated.  Our findings support this argument, \textcolor{black}{as does the low FPP computed from the $Kepler$ light curve} (Table \ref{fpp}).}

We note that the phase-folded $Kepler$ light curve given in the data validation summary from the NASA Exoplanet Archive for this target displays `brightening' features before ingress and after egress.  We find no evidence of this in our light curve, and we attribute this to be an artifact of the reduction process in the $Kepler$ pipeline.

This is another target that was noted by \cite{bn13} as being within the sub-Jovian desert, which we discuss further in \S\ref{pampas}.

\subsection{$Kepler$ False Positive Rate}
\label{fpr}

\textcolor{black}{In this paper we present observations that support the case for a planetary nature for KOI 439.01 and KOI 732.01 \textcolor{black}{and a likely false positive nature for KOI 531.01}.  Following the high false positive probabilities presented here and in \citet{morton14} and \citet{swift15}, we treat KOI 531.01 as a false positive in the discussion below.}  In Papers I and II, we identified a total of three false positives (KOI 565.01, 225.01, and 1187.01) and supported two candidates as validated planets (KOI 420.01 and 526.01).  Thus, of eight candidates observed in total, we identify four as viable planets and identify four as false positives.  This suggests a false positive rate as high as 50\% for short-period candidates.   \textcolor{black}{We note that KOI 565.01 and KOI 1187.01 were identified as false positives upon further analysis by the $Kepler$ team, so if we exclude these from our sample as well as KOI 531.01 we find that one of five or 20\% of short-period candidates are false positives.} 
\textcolor{black}{While our sample size is small, the higher false positive rate we estimate is consistent with results from \citet{fressin13} for short-period candidates.  Lower false positive rates of $\sim$10\% have been estimated for the entire sample of candidates (e.g. \citealp{morton11,morton12,fressin13,desert12}), while others have estimated false positive rates as high as $\sim$35\% for giant planets with $P$ $<$ 25 days \citep{santerne12} or for candidates due to some type of contamination \citep{coughlin14}.}  

In Figure \ref{perhist}, we present the latest distribution of $Kepler$ planet candidates, eclipsing binaries, and false positives.\footnote{Accessed from the NASA Exoplanet Archive (http://exoplanetarchive.ipac.caltech.edu/) on 2015 April 4 and the $Kepler$ eclipsing binary catalog (http://keplerebs.villanova.edu/) on 2014 June 25.}  The false positives are those KOIs that are listed with a $Kepler$ disposition of FALSE POSITIVE.  We note that the cumulative catalog was used to generate these distributions, which has not been fully vetted to date.  Therefore, some false positives could exist among the KOIs listed as candidates.  \textcolor{black}{Notably, KOI 225.01 is currently listed in the archive as a planet candidate, but it is clearly a false positive based on the significant wavelength-dependent transit depth (Paper II).  KOI 531.01 is also listed as a planet candidate, though it has a false positive probability $\apgt$50\% \citep{morton14,swift15}.}  These distributions clearly emphasize the continued prevalence of eclipsing binaries (and correspondingly, false positives) at the shortest orbital periods, as discussed in further detail in Paper II.  

Given the small sample size presented here, we cannot make strong statements regarding trends in the false positive rate with other planetary or stellar parameters.  Still, after investigating different parameters that included the $Kepler$ magnitude and Galactic latitude, we find no evidence of correlations in the properties of our identified false positives or their hosts.  However, we find that at short periods ($P \aplt 6$ days), the distribution of KOIs dispositioned as false positives is clearly skewed towards lower Galactic latitudes compared to the candidate population (Figure \ref{lathist}).  This is to be expected, given the increased density of stars closer to the Galactic plane.  Somewhat surprisingly, the distributions of $Kepler$-identified false positives and candidates in terms of their $Kepler$ magnitude are similar, except at the faintest magnitudes (Kp $>$ 16) where false positives are more prevalent (also as expected).

Lastly, we do not expect our findings to be otherwise biased by our selection criteria, though we note that our restriction on transit depth and/or duration could technically bias the spectral types we target.  We do not expect such a bias in our current target sample, given that the stellar effective temperatures of our targets range from about 4000 to 6000 K.

\begin{figure}
\includegraphics[scale=0.3, angle=90]{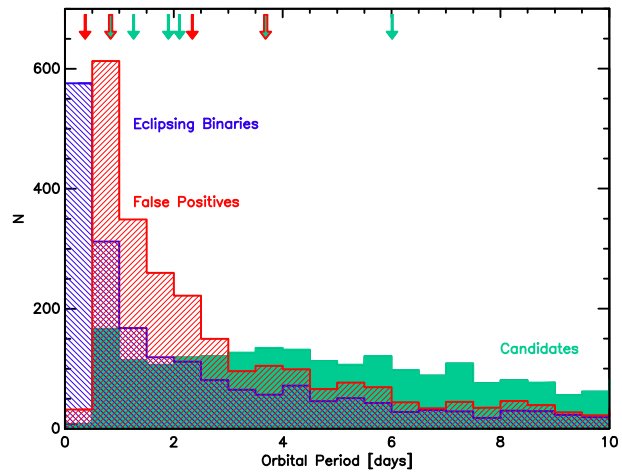}
\caption{Distribution of $Kepler$ planet candidates (green), false positives (red), and eclipsing binaries (blue) as a function of orbital period.  The arrows indicate the periods of the eight KOIs in our entire sample.  The green arrows indicate KOIs we tentatively identify as planets, and the red arrows indicate those KOIs identified as false positives.  The two red arrows filled with green are KOIs that are still listed in the $Kepler$ catalog as active planet candidates (\textcolor{black}{KOI 225.01 and KOI 531.01, though the false positive nature of the latter is not definitive}).}
\label{perhist}
\end{figure}

\begin{figure}
\includegraphics[scale=0.3, angle=90]{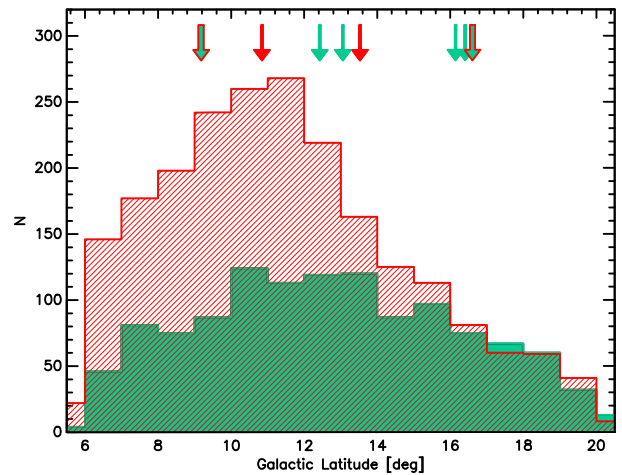}
\caption{Similar to Figure \ref{perhist}, but for the distribution of $Kepler$ planet candidates (green) and false positives (red) as a function of Galactic latitude.  Only planet candidates and false positives with periods less than 6.5 days are included.  The arrows are the same as in Figure \ref{perhist}.}
\label{lathist}
\end{figure}

\subsection{Planets in the sub-Jovian Desert}
\label{pampas}



With the increasing numbers of confirmed exoplanets and the thousands of planet candidates from $Kepler$, exciting trends have been emerging in the planet population.  In particular, there is a region in the period-radius plane (and correspondingly in the period-mass plane) where there appears to be no known planets.  This region has been referred to as the sub-Jovian desert and is the subject of several recent studies \citep[e.g.,][]{bl11,sk11,bn13,kurokawa14}.  The region encompasses orbital periods approximately less than 2.5 days and radii between about 3 and 11 $R_{\oplus}$ (Figure \ref{perrad}).  For comparison purposes, we show in Figure \ref{permass} the desert in the period-mass plane (where it spans approximately 0.03$-$1 $M_J$).  

\textcolor{black}{We note that in Figs \ref{perrad} and \ref{permass}, the desert region is overemphasized because we show planets discovered by both radial velocity surveys and ground- and space-based transit surveys.  This combined sample therefore suffers from many biases.  Still,} the lack of planets in this region cannot be easily explained by observational bias.  Short-period planets are generally easier to detect, especially if they are larger or more massive than Earth.  Assuming the desert is not due to observational bias, this suggests that there is some formation mechanism at play.  For instance, if orbital migration is required for planets to end up on close-in orbits, then there might be a limiting mass where planets can no longer migrate inward.  The pile-up of hot Jupiters around $\sim$3 days is suggestive of this.  However, since smaller/less massive planets are (apparently) allowed to orbit very close to their host stars (e.g. with periods less than 1 day), it has been theorized that these small planets are the leftover cores of planets that had their atmospheres blown away by the host star (e.g. \citealp{youdin11}).  This mechanism is likely most effective for planets with low surface gravity, and not necessarily for all hot Jupiters.  \citet{owen13} investigated evaporation in detail and conclude that evaporation explains a lack of low density planets on close-in orbits as well as a deficit of planets around 2 $R_{\oplus}$.  Currently, this seems to be the most viable explanation to explain the observed distribution of planets and candidates.  While beyond the scope of this paper, \citet{bn13} offer a more complete discussion of alternative causes of the desert, including that hot Jupiters may have been tidally captured while super-Earth-size planets may have been affected by disk-planet interactions.  

Given the competing formation mechanisms, any candidate planets within this desert are therefore of extreme interest. \textcolor{black}{In total, there are currently $\sim$30 KOIs within the desert, including KOI 439.01 and KOI 732.01 which we identify as likely planets here.}  However, several KOIs have properties that place them near the boundaries of the region, and uncertainties in the planetary radius can easily shift some of them out of the desert.  Follow-up observations of these KOIs are needed to confirm their planetary nature.

At least three of the KOIs in the desert are in known multiple candidate systems, suggesting that these are likely real planets.  \textcolor{black}{Along with our observations providing additional evidence in support of KOI 439.01 and 732.01 as viable planets}, and three other confirmed planets (Kepler-41b, Kepler-119b, WASP-43b),\footnote{Determined from parameters given on exoplanets.org} there could be at least eight known planets located in the desert so far (albeit towards the edges of the region).  It will be interesting to see if the KOIs that are in the more central region of the desert survive the vetting process, considering the \textcolor{black}{higher false positive rate that exists for KOIs with short periods compared to the global rate}.  It may be that planets can exist in this regime, but under atypical circumstances of planetary formation and migration.  If none do, it at least appears that some planets are filling out the edges of the region and as a result, the boundaries of the desert are being constrained.  The continued study of planets in this region is therefore critical for informing planetary formation theories and at the very least will help to constrain the boundaries of the desert.

\begin{figure}
\includegraphics[scale=0.3, angle=90]{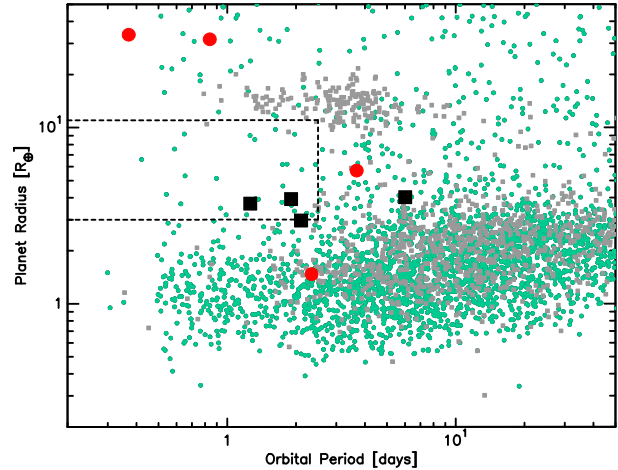}
\caption{\textcolor{black}{Planet radius versus orbital period for all confirmed transiting planets (from both ground-based surveys and $Kepler$; gray squares)} and for active $Kepler$ planet candidates (green circles).  The black squares mark the locations of the four KOIs we \textcolor{black}{support} as planets, while the red circles mark the four KOIs that we identified as false positives. \textcolor{black}{We consider KOI 439.01 and 732.01 to be planets and KOI 531.01 to be a false positive in this context.}  The parameters for confirmed planets were taken from The Exoplanet Orbit Database at exoplanets.org on 2014 July 31.  The region marked by dashed black lines indicates the regime of the so-called sub-Jovian desert (for $R_p$ $\sim$ 3$-$11 $R_{\oplus}$ and $P$ $<$ 2.5 days).  \textcolor{black}{While we initially selected KOIs with radii less than 6 $R_{\oplus}$, updated stellar parameters from \citet{huber14} and light curve models from the $Kepler$ team yielded new planet radii of 33.63 and 31.69 $R_{\oplus}$ for the false positives KOI 1187.01 and 225.01 (the two shortest period KOIs we observed, the latter of which is notably still included in the archive as a planet candidate).}}
\label{perrad}
\end{figure}

\begin{figure}
\includegraphics[scale=0.3, angle=90]{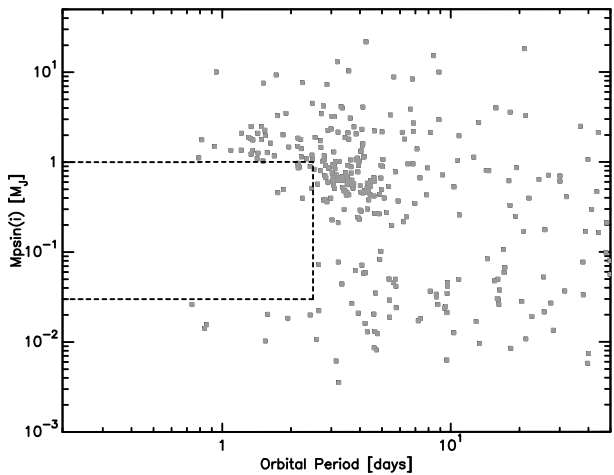}
\caption{\textcolor{black}{Similar to Figure \ref{perrad}, but showing the minimum planet mass versus orbital period for all confirmed planets (from both radial velocity and transit surveys)}.  The region marked by dashed black lines covers the mass range of 0.03$-$1 $M_J$ and $P$ $<$ 2.5 days.}
\label{permass}
\end{figure}

\section{Conclusion}
\label{sconc}
\textcolor{black}{In this paper, we use multi-wavelength transit photometry to argue for a planetary nature for the two $Kepler$ planet candidates KOI 439.01 and KOI 732.01 and a false positive identification for KOI 531.01.}  \textcolor{black}{Additional follow-up observations, such as radial velocity and/or photometry at redder wavelengths, are needed to clarify the nature of KOI 439.01 in particular.}  In combination with results from previously published work \citep{colon11,colon12}, we have vetted a total of eight small, short-period $Kepler$ candidates through GTC observations.  While our sample size is small, our results support a slightly higher false positive rate for short-period planet candidates than for the overall sample.  This is in agreement with results from other studies and} has interesting implications for $Kepler$ planet candidates located within the sub-Jovian desert, a region in the period-radius plane where some planetary formation mechanisms appear to restrict planets from living there.  Future observations of these candidates will improve our understanding of different formation scenarios, particularly if none of these candidates survive the vetting process.  

\section*{Acknowledgments }

We acknowledge the anonymous referee for helping us greatly improve this paper.  This research is based on observations made with the Gran Telescopio Canarias (GTC), installed in the Spanish Observatorio del Roque de los Muchachos of the Instituto de Astrof\'isica de Canarias, in the island of La Palma.  We gratefully acknowledge the GTC observing staff for helping us plan and conduct these observations.  We acknowledge the University of Florida College of Liberal Arts and Sciences and the Department of Astronomy, where a portion of this work was completed.  We also acknowledge the $Kepler$ Eclipsing Binary Working Group for their assistance in better understanding the contents of the eclipsing binary catalog, especially Joshua Pepper, Kyle Conroy, Jeff Coughlin, Dave Latham, and Andrej Prsa.  K.D.C. acknowledges previous support and R.C.M. acknowledges current support from the National Science Foundation under Grant No. DGE1255832.  Any opinions, findings, and conclusions or recommendations expressed in this material are those of the author(s) and do not necessarily reflect the views of the National Science Foundation.  E.B.F. acknowledges support by the National Aeronautics and Space Administration under grants NNX12AF73G and NNX14AN76G issued through the $Kepler$ Participating Scientist Program.  This research has made use of the NASA Exoplanet Archive, which is operated by the California Institute of Technology, under contract with the National Aeronautics and Space Administration under the Exoplanet Exploration Program.  This research has also made use of the Exoplanet Orbit Database and the Exoplanet Data Explorer at exoplanets.org.

\label{lastpage}

\end{document}